\documentclass[%
reprint,
nofootinbib,
nobibnotes,
amsmath,amssymb,
aps,
pre,
]{revtex4-1}
\usepackage{graphicx}
\usepackage{bm}
\usepackage{color}

\newcommand{\xib}{\bar{\xi}}
\newcommand{\etab}{\bar{\eta}}

\newcommand{\epsr}{\epsilon}

\newcommand{\unitz}{\hat{\bm{z}}}
\def\d{\mathrm{d}}
\def\b#1{\mathbf{#1}}

\begin{document}
\title{Laplace's equation for a point source near a sphere: improved  internal solution using spheroidal harmonics}

\author{Matt R. A. Maji\'c} 
\author{Baptiste Augui\'e}
\author{Eric C. Le Ru} \email{eric.leru@vuw.ac.nz}

\affiliation{The MacDiarmid Institute for Advanced Materials and Nanotechnology,
School of Chemical and Physical Sciences, Victoria University of Wellington,
PO Box 600, Wellington 6140, New Zealand}

\date{\today}

\begin{abstract}
As shown recently [Phys. Rev. E 95, 033307 (2017)], spheroidal harmonics expansions are well suited for the external solution of Laplace's equation for a point source outside a spherical object. Their intrinsic singularity matches the line singularity of the analytic continuation of the solution and the series solution converges much faster than the standard spherical harmonic solution. Here we extend this approach to internal potentials using the Kelvin transformation, ie. radial inversion, of the spheroidal coordinate system. This transform converts the standard series solution involving regular solid spherical harmonics into a series of irregular spherical harmonics. We then substitute the expansion of irregular spherical harmonics in terms of transformed irregular spheroidal harmonics into the potential. The spheroidal harmonic solution fits the image line singularity of the solution exactly and converges much faster. We also discuss why a solution in terms of regular solid spheroidal harmonics cannot work, even though these functions are finite everywhere in the sphere. We also present the analogous solution for an internal point source, and two new relationships between the solid spherical and spheroidal harmonics.
\end{abstract}

\maketitle

\section{Introduction}

In \cite{2017SuperLaplace} we expressed the solution of Laplace's equation for a point source outside a sphere as series of solid spheroidal harmonics, which converges much faster than the standard spherical harmonic series. This is intriguing as the spherical coordinate system initially appeared as more natural.
Here we consider the solution inside the sphere for the same problem of a point charge near a dielectric sphere. The conclusions
can be easily extended to any multipolar point source and to other physical systems governed by Laplace's equation.

The method we previously used to express the external potential in terms of spheroidal harmonics is to substitute the expansion of irregular spherical harmonics in terms of irregular spheroidal harmonics into the standard series solution and rearrange the order of summation \cite{2017SuperLaplace} (although the spheroidal solution can be derived on its own using standard methods, but with more effort). For internal potentials, the standard solution is a series of regular solid spherical harmonics, so it would seem natural to substitute the expansion of regular solid spherical harmonics in terms of regular solid spheroidal harmonics. To this end, we present and prove (see Appendix) two relationships between the spherical and offset spheroidal harmonics, similar to the ones presented in \cite{2017SuperLaplace}.
However, as discussed in detail in section \ref{regExpansion}, this approach fails to produce a suitable solution. 
Instead, as presented in Sec. \ref{SecInternal}, the spherical solution can be manipulated by radial inversion ($r\rightarrow1/r$), as first discussed by Kelvin in 1845 \cite{1847Thomson}
and often known as the Kelvin transformation. This transform is a conformal mapping often used to transform the geometry of a complex problem into a simpler geometry in which the solution is known \cite{2009Dassios,2017Kelvin}. We apply the transform to turn the sphere inside out, so that the problem becomes equivalent to finding the external potential in the transformed frame. We can then follow the same method as in \cite{2017SuperLaplace} to obtain the solution as a series of transformed irregular spheroidal harmonics. Again the new solution converges much faster than the standard spherical harmonics solution. The similar case of an internal source is treated  in Sec. \ref{SecInside}.

The problem has been solved in the past using spherical harmonic series \cite{1941Stratton} or using the method of images, first reported by Neumann \cite{1883Neumann} and revisited more recently by Poladian \cite{poladian1988} and Lindell \cite{1992Lindell}\footnote{Note that in \cite{2017SuperLaplace}, the first solution was wrongly attributed to Lindell.}.
The integral solution is the analytic continuation of the potential, which consists of an integral over a line. This line singularity of is naturally interpreted as an image source, which could not identified from the spherical harmonic series. Here we show that the singularities of the inverted spheroidal harmonics lie exactly on the image line charge. This work further highlights the importance of using basis functions with singularities matching that of the solution, and makes this connection more precisely using the Havelock formula \cite{1952Havelock,1974Miloh}. 
This work also leads us to introduce an uncommon type of coordinates: radially-inverted offset prolate spheroidal coordinates, a partially separable coordinate system of the Laplacian.
Similar radially-inverted coordinate systems are discussed by Moon and Spencer \cite{1961FieldTheoryHandbook} and used in acoustic scattering \cite{dassios1999rayleigh}, and fluid dynamics \cite{2015InvertedSpheroidal}, although in these texts the spheroidal coordinates are centred about the origin.

\begin{figure}
\includegraphics[width=7cm,trim={4.8cm 16cm 6cm 3cm},clip]{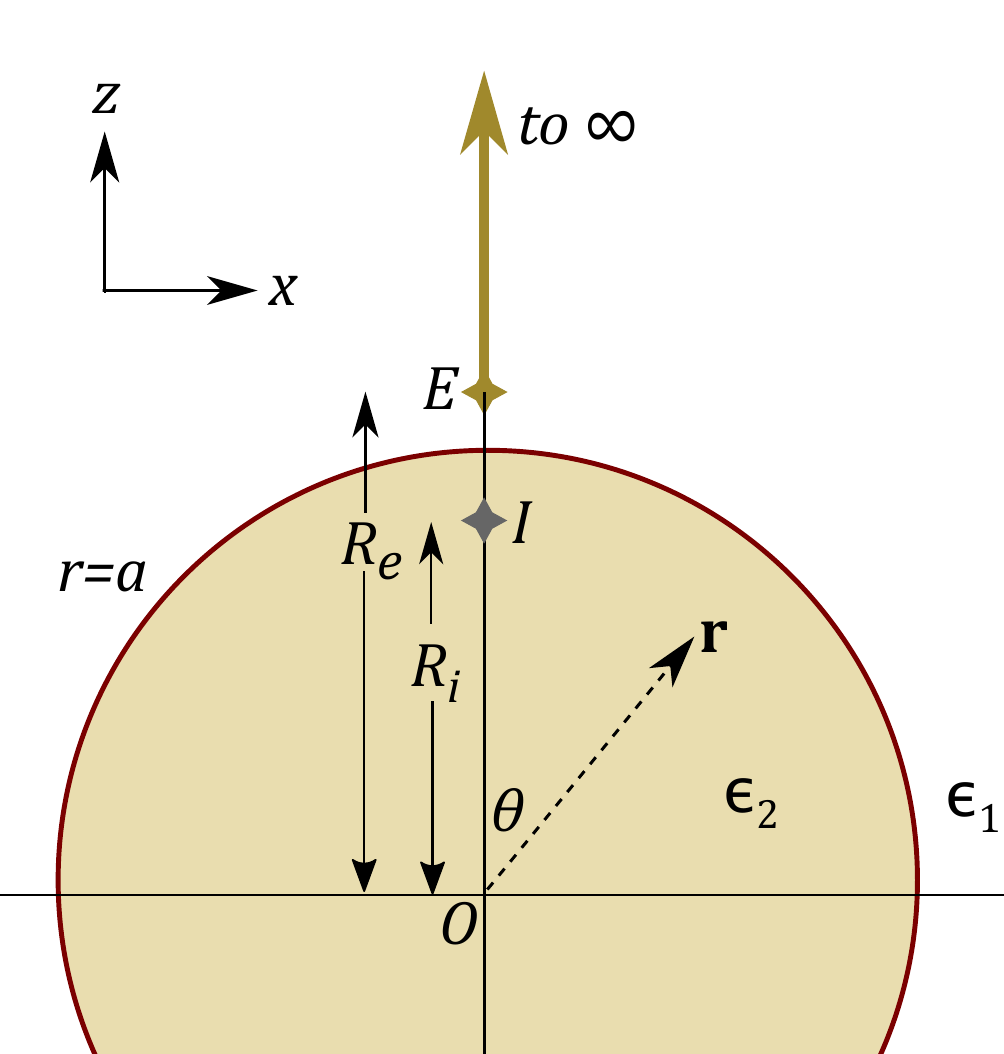}
\caption{Schematic of the problem considered in this work. A point charge $q$ is located at point E (external) outside a dielectric sphere of radius $a$, on the $z$-axis a distance $R_e$ from the origin. Point I (internal) located at $z=R_i=a^2/R_e$ is the location of the image charge in the standard solution for the outside potential. The relative permittivity of the sphere with respect to the surroundings is $\epsilon=\epsilon_2/\epsilon_1$. The line from E to infinity is the location of the image line charge used in the solution of the internal potential within the method of images \cite{1992Lindell}. 
The external solution was studied in Ref.~\cite{2017SuperLaplace} and we focus on the internal solution in Secs.~\ref{SecInternal} and \ref{regExpansion}. Sec. \ref{SecInside} considers the case where the point charge $q$ is located inside the sphere at point I.} \label{FigSchem}
\end{figure}

\section{Point charge outside the sphere}
\label{SecInternal}

We consider the same problem as in Ref. \cite{2017SuperLaplace}, shown schematically in Fig. \ref{FigSchem}.
For convenience, we write the potential as $V = \bar{V}\cdot q/(4\pi\epsilon_0\epsilon_1 a)$ and work with the dimensionless $\bar{V}$. The potential outside the sphere is $\bar{V}_\mathrm{out} = \bar{V}_q + \bar{V}_r$, where $\bar{V}_q$ is the potential of the point charge and $\bar{V}_r$ the reflected potential.
$\bar{V}_\mathrm{in}$ is the potential inside the sphere.
The standard solution in spherical coordinates $(r,\theta,\phi)$ is  \cite{1941Stratton,1992Lindell}:
\begin{align}
\bar{V}_q&= \frac{a}{|{\bf r} - R_e \unitz|}=\frac{a}{R_e} \sum_{n=0}^\infty \left(\frac{r}{R_e}\right)^n P_n(\cos\theta)  \qquad (r<R_e), \label{EqVqSph}\\
\bar{V}_r&= - \sum_{n=0}^\infty \beta_n \left(\frac{R_i}{r}\right)^{n+1} P_n(\cos\theta)  \qquad (r\ge a), \label{EqVrSph}\\
\bar{V}_\mathrm{in}&= \frac{a}{R_e} \sum_{n=0}^\infty \frac{2n+1}{n(\epsilon+1)+1}\left(\frac{r}{R_e}\right)^{n} P_n(\cos\theta)  \qquad (r\le a),
\label{EqVinSph}
\end{align}
where $R_i=a^2/R_e$, $P_n$ are the Legendre polynomials, and
\begin{align}
\beta_n = \frac{n(\epsilon-1)}{n(\epsilon+1)+1},\quad \text{with }
\beta_\infty = \frac{\epsilon-1}{\epsilon+1}.
\end{align}
for a source close to the surface, $R_e\rightarrow a$, and for $r\rightarrow a$, these series are slowly convergent.

\subsection{External potential}
In Ref.~\cite{2017SuperLaplace}, the irregular spherical harmonics in Eq.~\ref{EqVrSph} were expanded in terms of irregular spheroidal harmonics $Q_n(\bar{\xi})P_n(\bar{\eta})$ where $(\bar{\xi},\bar{\eta},\phi)$ are spheroidal coordinates with foci at O and I (see Eq.~\ref{PnmvsQPnm}), and $Q_n$ denotes the Legendre functions of the second kind. The resulting expression is
\begin{align}
\bar{V}_r =
 -\beta_\infty \frac{R_i}{r'} + 2\beta_\infty \sum_{n=0}^\infty ~ (2n+1) c_n Q_n(\xib) P_n(\etab). 
\label{EqnPhiSca6}
\end{align}
where
\begin{align}
c_n &=\mu\sum_{k=0}^n\frac{(n+k)!}{(n-k)!k!^2}\frac{(-)^{n+k}}{k+\mu} \nonumber\\
&=\prod_{k=0}^{n} \frac{\mu-k}{\mu+k}, \qquad \mu = \frac{1}{\epsr+1}.
\label{EqnCn}
\end{align}
The first term on the right corresponds to an image charge, with $r'=\sqrt{r^2-2R_ir\cos\theta+R_i^2}$. Note that $Q_n$ should be computed with a backward recurrence scheme as outlined in \cite{2017SuperLaplace}.

\begin{figure*}[t]
\includegraphics[width=\textwidth,clip=true,trim=0.5cm 12cm 1.5cm 0cm]{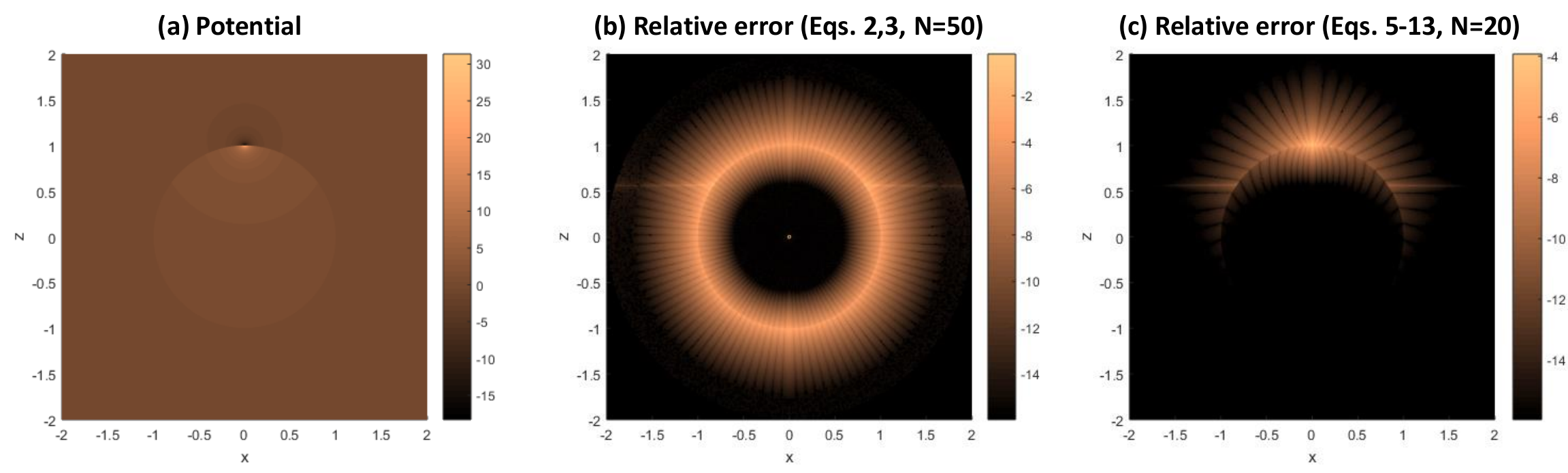}
\caption{Surface maps of the solution for a point charge located at $z=R_e=1.02 a$ just outside a sphere of radius $a=1$ and relative dielectric constant $\epsilon=2.25$.
(a) Plots of the converged solutions for the inside potential $\bar{V}_\mathrm{in}$ and reflected potential $\bar{V}_{r}$, computed from Eqs. 5 and 13 with $N=130$ terms in the series.
(b-c) Corresponding plots of the relative error (in $\log_{10}$ scale) using the spherical harmonic series with $N=50$ (b) or the spheroidal harmonic series with $N=20$ (c).
The horizontal line around $z=0.5$ is due to the reflected potential being very small here, which makes the relative error large. The dark radial lines in (b-c) correspond to regions where the approximate solution
coincidentally crosses the exact solution because it oscillates spatially around the correct value. 
} \label{errplots}
\end{figure*}

\subsection{Internal potential}

We now apply the same approach to the internal potential. Following \cite{2017SuperLaplace}, we first separate the dominant term in the series coefficients:
\begin{align}
\frac{2n+1}{n(\epsilon+1)+1} &= \frac{2}{\epsr+1} + \frac{\beta_\infty}{n(\epsr+1)+1},
\end{align}
and recognize the sum over the constant term as $\bar{V_q}$:
\begin{align}
\bar{V}_\mathrm{in}&=\frac{2\bar{V}_q}{\epsr+1} + \frac{a}{R_e} \sum_{n=0}^\infty \frac{\beta_\infty}{n(\epsilon+1)+1}\left(\frac{r}{R_e}\right)^n P_n(\cos\theta). \label{EqVin2}
\end{align}

From here the natural approach would be to expand the regular spherical harmonics in terms of regular spheroidal harmonics. However, for reasons discussed in the next section, this does not work.

Instead we use the Kelvin transformation to transform the expression into one that is very similar to that of the reflected potential.
The Kelvin transformation is an inversion of the radial coordinate and is defined in \cite{2009Dassios,2017Kelvin}:
\begin{align}
\check{r} = \frac{a^2}{r}, \qquad \check{R}_e=\frac{a^2}{R_e}=R_i,\label{radial_inv} 
\end{align}
and $\theta, \phi$ are unchanged. With the transformed coordinates, Eq.~\ref{EqVin2} becomes (for $\check{r} \ge a$)
\begin{align}
\bar{V}_\mathrm{in}&=\frac{2\bar{V}_q}{\epsr+1} + \frac{a}{r}\sum_{n=0}^\infty\frac{\beta_\infty}{n(\epsilon+1)+1}\left(\frac{\check{R}_e}{\check{r}}\right)^{n+1} P_n(\cos\theta) \label{EqVin3}
\end{align}
We can therefore apply the expansion of irregular solid spherical harmonics of ($\check{r},\theta,\phi$) in terms of offset irregular spheroidal harmonics (see Eq.~\ref{PnmvsQPnm}), which can be written as:
\begin{align}
&{\left(\frac{\check{R}_e}{\check{r}}\right)^{n+1}} P_n(\cos\theta) =
 \nonumber\\
& \sum_{k=n}^\infty (-1)^{n+k} \frac{2(2k+1)(k+n)!}{n!^2(k-n)!} ~ Q_k(\check\xi) P_k(\check\eta). \label{PvsQPgrave}
\end{align}
The ``radially-inverted offset spheroidal coordinates'' $\check\xi$, $\check\eta$ are:
\begin{align}
\check\xi&=\frac{\check{r}+\check{r}'}{\check{R}_e}, \qquad
\check\eta=\frac{\check{r}-\check{r}'}{\check{R}_e},\\
\mathrm{where~}\check{r}' &= \sqrt{\check{r}^2-2\check{r}\check{R}_e\cos\theta+\check{R}_e^2}. \nonumber
\end{align}
Their domains are $\check{\xi}\in[1,\infty)$, $\check{\eta}\in[-1,1]$, the same as for non-inverted spheroidal coordinates. 
Eq.~\ref{PvsQPgrave} is valid everywhere except for $\check{\xi}=1$ or equivalently $\theta=0,~~0 \le \check{r} \le \check{R_e}$, which corresponds to the infinite segment on the positive $z$-axis with $z> R_e$ (outside the sphere).
In general, if $f(r,\theta,\phi)$ satisfies Laplace's equation, then $f(a^2/r,\theta,\phi)/r$ is also a solution, so $Q_n(\check\xi)P_n(\check\eta)/r$ must be a solution. Since the non-inverted irregular spheroidal harmonics are suitable for modelling potentials that go to zero at infinity, the inverted harmonics should be suitable for potentials that are finite at the origin. The corresponding iso-potentials of $Q_n(\check\xi)P_n(\check\eta)/r$ are shown in Appendix \ref{SecInvSph}.

Following Ref.~\cite{2017SuperLaplace}, we then substitute the expansion in Eq. \ref{PvsQPgrave} into Eq. \ref{EqVin3}, change the order of summation, re-label $k\leftrightarrow n$ and use Eq.~\ref{EqnCn} to obtain
\begin{align}
\bar{V}_\mathrm{in}\!=\frac{2\bar{V}_q}{\epsr+1} + 2\beta_\infty\frac{a}{r}
\sum_{n=0}^\infty(2n\!+\!1)c_n~ Q_n(\check\xi)P_n(\check\eta), \label{EqVin4}
\end{align}
which is valid everywhere inside the sphere.
As was found for the external potential, Eq.~\ref{EqVin4} converges much faster than the standard solution as shown in Figs.~\ref{conv}-\ref{errplots}. This is related to the singularity of the solution. The potential can be represented in integral form as:
\begin{align}
\bar{V}_\mathrm{in} = \frac{2\bar{V}_q}{\epsr+1} + \frac{a}{R_e}\frac{\epsilon\beta_\infty}{\epsr+1} 
\int_{R_e}^\infty\frac{(R_e/\tilde{z})^\mu~ \d \tilde{z}}{\sqrt{x^2+y^2+(z-\tilde{z})^2}}.
\end{align}
The last term is the the potential created by an infinite line charge on the $z$ axis for $z \geq R_e$. This corresponds exactly to $\check{\xi}=1$, i.e. to the singularity of $Q_k(\check\xi)$, which makes the inverted spheroidal harmonics an ideal basis for the problem. 

We would like to point out that although the integral solution can be evaluated numerically to any degree of accuracy, series solutions have advantages. For example, series convergence can be more easily tested than integral quadrature accuracies. A series solution also lends itself more easily to further analytic derivations of other physical quantities (for example the flux of the field over a closed surface).

\begin{figure}
\includegraphics[width=7cm,trim={4.6cm 13cm 6.4cm 5.1cm},clip]{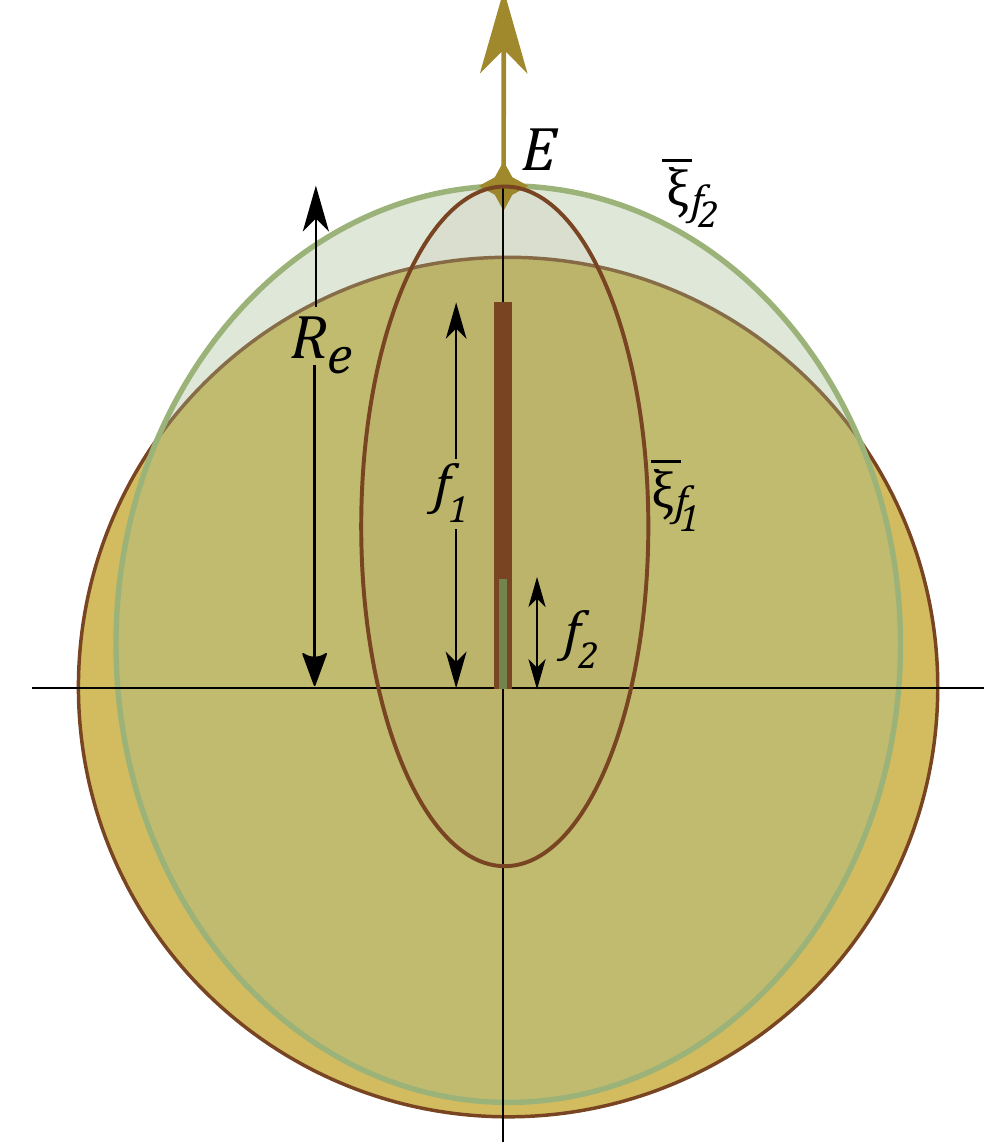}
\caption{Boundary of convergence for the series solution expressed on two different bases of offset regular spheroidal harmonics of focal lengths $f_1$ and $f_2$. Any series of regular spheroidal harmonics will diverge outside some spheroid and converge (to a finite value) inside. The spheroid must be small enough so that it excludes the singularity of the analytic continuation of the series, which extends from point E $z=R_e$ for the internal potential. A choice of a larger focal length will mean that the spheroid boundary is too thin and the series will diverge for most of the interior of the sphere. If the focal length is very small, the boundary of convergence approaches a sphere, the same spherical boundary that defines the convergence of the standard spherical solution, but the spheroidal harmonics also reduce to spherical harmonics.} \label{spheroids}
\end{figure}

\section{Problems with solutions in terms of regular spheroidal harmonics} \label{regExpansion}

We now analyze why it is impractical to expand the internal potential on a basis of regular spheroidal harmonics. The regular spherical harmonics in Eq.~\ref{EqVin2} can be expanded in terms of offset regular spheroidal harmonics just as can be done for the irregular harmonics; in fact the
expansion is finite:
\begin{align}
\left(\frac{r}{R_e}\right)^n\! P_n(\cos\theta)=\sum_{k=0}^n\frac{n!^2(2k+1)}{(n-k)!(n+k+1)!} P_k(\bar\xi_e)P_k(\bar\eta_e) \label{PvsPP}.
\end{align}
Eq. \ref{PvsPP} appears new and is proved in the Appendix. Note that the offset spheroidal coordinates ($\xib_e$,$\etab_e$) are here defined with O and E as foci, not O and I as used for the reflected potential.
The expansion coefficients of the internal potential in terms of regular spheroidal harmonics can be found by substituting Eq. \ref{PvsPP} into the potential given in Eq.~\ref{EqVin2}, rearranging the order of summation and relabelling $n\leftrightarrow k$:
\begin{align}
\bar{V}_\mathrm{in} =\frac{2\bar{V}_q}{\epsr+1} + \frac{a \beta_\infty}{R_e} \sum_{n=0}^\infty(2n+1) d_n P_n(\bar\xi_e)P_n(\bar\eta_e), \label{VPP}\\
\mathrm{where~} d_n = \sum_{k=n}^\infty\frac{1}{k(\epsilon+1)+1} \frac{k!^2}{(k-n)!(k+n+1)!}. \nonumber 
\end{align}
The sum over $k$ in the definition of $d_n$ converges, but we could not find a simple closed form of these coefficients as was the case for $c_n$ in the solution of the reflected potential (Eq.~\ref{EqnCn}).

\begin{figure*}[t!]
\center
\includegraphics[scale=.6]{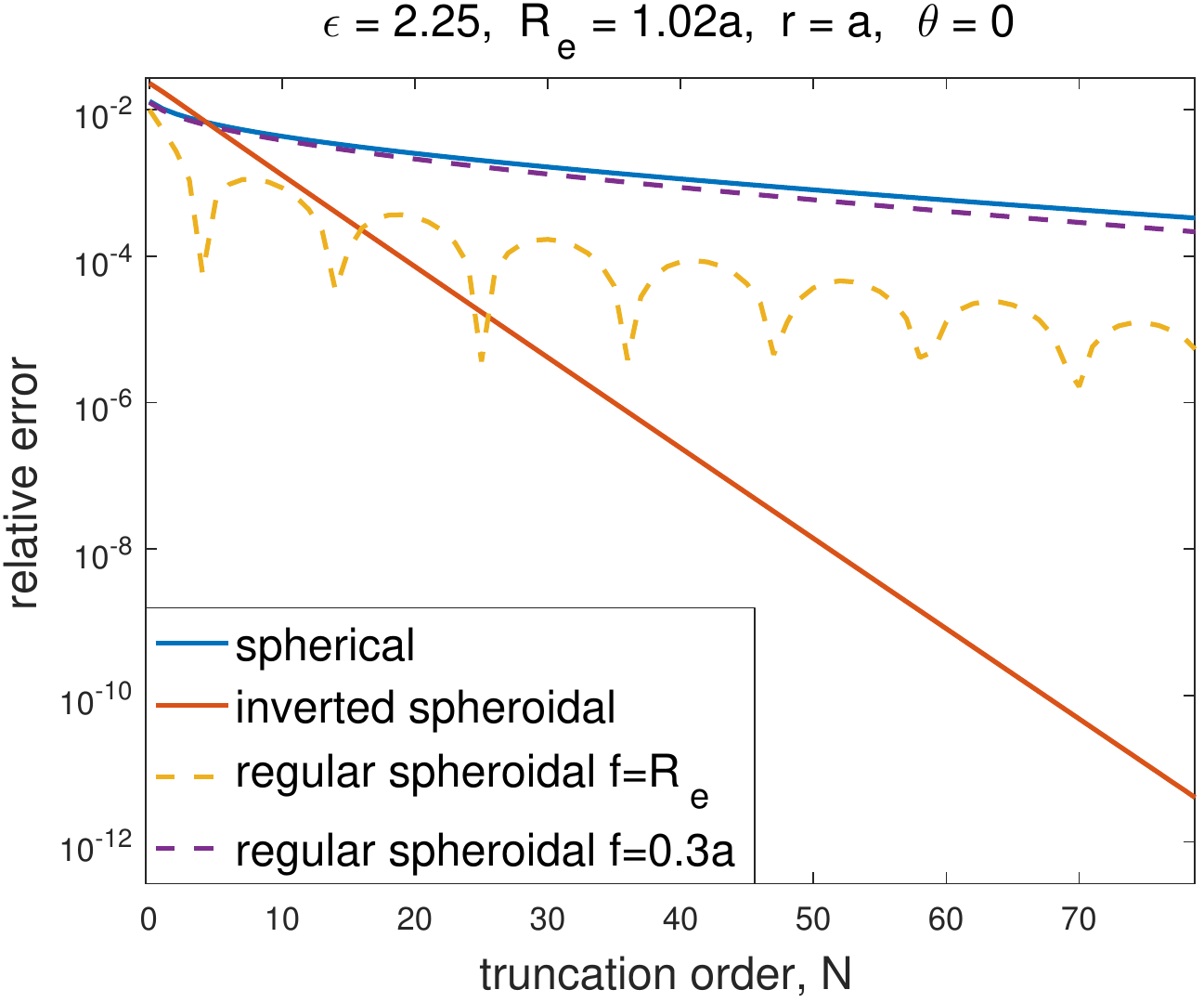} 
\includegraphics[scale=.6]{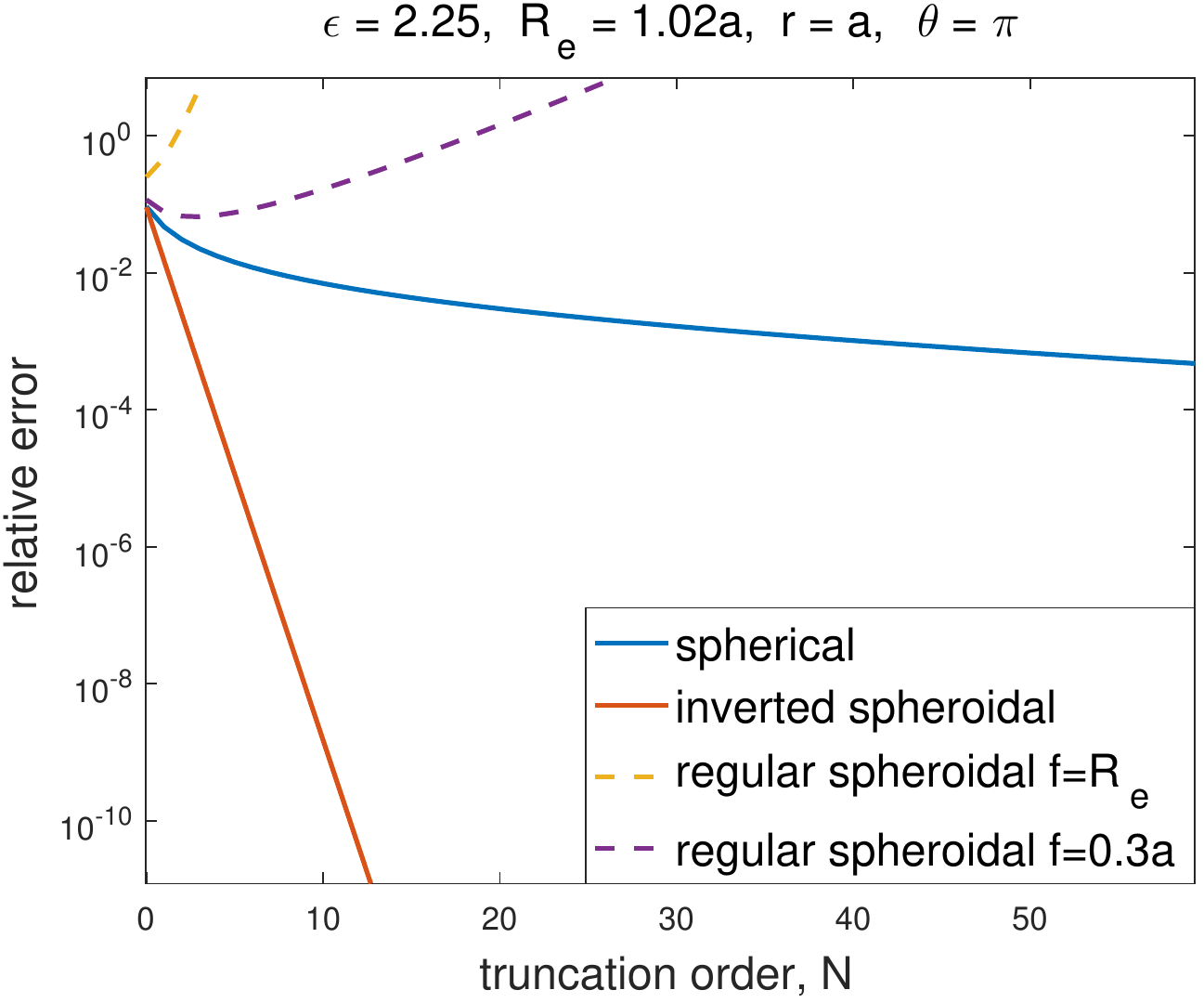}
\caption{Comparison of the convergence rates of the internal potential series  using spherical (Eq.~\ref{EqVinSph}), inverted spheroidal (Eq.~\ref{EqVin4}), and regular spheroidal harmonics with $f=R_e$ (Eq.~\ref{VPP}) and $f=0.3a$ (Eq.~\ref{VPPF}). This is quantified as the relative error between the $N^{th}$ partial sum and the ``converged'' result. The source is at $0.02 a$ from a sphere with $\epsilon=2.25$. The results are qualitatively $\epsilon$-independent, except for extreme values such as for perfect conductors or $\epsilon=-1$. We compute the internal potential at two opposite points, $\theta=0$ (left) and $\theta=\pi$ (right) on the sphere surface $r=a$ (where the convergence is slowest).}
\label{conv}
\end{figure*}

The problem in Eq.~\ref{VPP} lies with the region of convergence of the series (of $n$). One can expect that the boundary of divergence will be the largest spheroid (constant $\bar\xi_e$) which does not cross the singularity of the analytic continuation of the series (see App. \ref{convProof} for proof). But the singularity extends from point E to infinity and any spheroid with $\bar\xi_e>1$ crosses this line. Consequently the series only converges for $\xib_e=1$ -- on the $z$ axis for $0\leq z\leq R_e$ -- ideally we need it for all $r\le a$. 
As illustrated in Fig.~\ref{conv}, while there is a small improvement in the convergence rate of this series at $r=a,\theta=0$, the series diverges on the other side at $r=a,\theta=\pi$.

To avoid this problem, we search for a solution using offset spheroidal coordinates ($\bar\xi_f, \bar\eta_f$) centred on O and F=$(0,0,f)$, with a smaller focal length $f$, as shown in Fig.~\ref{spheroids}. Writing $(r/R_e)^n = (r/f)^n (f/R_e)^n$
and using Eq. \ref{PvsPP} substituting $f$ for $R_e$, we obtain after the same manipulations: 
\begin{align}
\bar{V}_\mathrm{in} =\frac{2\bar{V}_q}{\epsr+1} + \frac{a \beta_\infty}{R_e} \sum_{n=0}^\infty(2n+1) e_n P_n(\bar\xi_f)P_n(\bar\eta_f), \label{VPPF}\\
\mathrm{where~} e_n = \sum_{k=n}^\infty\frac{(f/R_e)^k}{k(\epsilon+1)+1} \frac{k!^2}{(k-n)!(k+n+1)!}. \nonumber 
\end{align}
The region of convergence of this series is again bounded by the spheroid surface (constant $\bar\xi_f$) that touches the base of the image singularity, i.e. it is defined by $\xib_f \le 2R_e/f-1$.
For a smaller $f$, this region can be larger than before where we had $f=R_e$, see Fig~\ref{spheroids}. By choosing $f \le R_e - a$, the region of convergence even contains the entire sphere $r\le a$, as desired. However, for a point source close to the sphere, we then have $f \ll a$, i.e. F approaches O, and the spheroidal harmonics will then approach the spherical harmonics (up to some normalization), so the series convergence becomes similar to the standard solution in spherical coordinates. Besides, this spheroidal harmonic series is impractical as the coefficients $e_n$ are also defined as a slowly-converging infinite sum.

Finally, we note that we could have used the regular spheroidal harmonics centred about the origin instead of the offset ones, but it does not remove the issues discussed above. Overall, all attempts at finding a solution in terms of regular spheroidal harmonics result in either series that do not converge over the entire interior sphere or in series that converge everywhere inside, but at a rate comparable with the original spherical harmonic solution.


\section{Point charge inside sphere}
\label{SecInside} 

The problem is almost the same as in Fig.~\ref{FigSchem} but now with the source located at point I ($z=R_i$), and the image charge at E.
The potential inside the sphere ($r\le a$) is written as $\bar{V}_\mathrm{in} = \bar{V}_q + \bar{V}_r$, where $\bar{V}_q$ is the potential due to the point charge and $\bar{V}_r$ the reflected potential. We now expand $\bar{V}_q$ on a series of irregular solid spherical harmonics centred at the origin:
\begin{align}
\bar{V}_q&=\frac{a}{|{\b r} - R_i\unitz|} = \frac{a}{R_i}\sum_{n=0}^\infty \left(\frac{R_i}{r}\right)^{n+1} P_n(\cos\theta)  \qquad (r>R_i) \label{pointchg}
\end{align}
The standard solution is
\begin{align}
\bar{V}_r&=\sum_{n=0}^\infty\frac{(n+1)(\epsilon-1)}{n(\epsilon+1)+1}
\left(\frac{r}{R_e}\right)^n P_n(\cos\theta), \label{pot1} \\
\bar{V}_\mathrm{out}&=\frac{a}{R_i} \sum_{n=0}^\infty \frac{2n+1}{n(\epsilon+1)+1}
\left(\frac{R_i}{r}\right)^{n+1} P_n(\cos\theta).
\end{align}
Following similar derivations to the problem for a point source outside the sphere, we find
\begin{align}
\bar{V}_\text{r}&=\frac{\beta_\infty R_e}{|{\b r} - R_e\unitz|} + 2\beta_\infty \epsr\frac{R_e}{r}
\sum_{n=0}^\infty(2n\!+\!1)c_n~ Q_n(\check\xi)P_n(\check\eta). \\
\bar{V}_\mathrm{out}&=\frac{2\bar{V}_q}{\epsr+1}  + 2\frac{a\beta_\infty}{R_i} 
\sum_{n=0}^\infty(2n\!+\!1)c_n~ Q_n(\xib)P_n(\etab).
\end{align}

\section{Discussion and conclusion}

This work further demonstrates that spheroidal harmonics provide a more suitable basis for the solution of Laplace's equation for a point source near a sphere.
Despite the irregular solid spheroidal harmonics being an ideal basis for the external potential, regular spheroidal harmonics are {\it not} a suitable basis for the internal potential.
Instead, the Kelvin transformation can be used to find solution in terms of radially-inverted offset spheroidal coordinates.  

The internal potential solution presented here further highlights the connection with solutions in terms of the method of images \cite{poladian1988,1992Lindell,1992LindellInside},
which was hinted at in Ref.~\cite{2017SuperLaplace}.
The solution in a given region of space has a unique analytic continuation with well-defined singularities, independently of the series expansions chosen to calculate it.
The best (most rapidly convergent) basis functions for the problem are likely those with singularities that match the singularity of the solution. 
In the case of the external potential for an external point source at E, the singularity is a line between O and I, and spheroidal harmonics of the
offset coordinates $\xib,\etab$ exactly match this.
In the case of the internal potential for an external point source at E, the singularity is a line extending from the point source at E to infinity
and the radially-inverted offset spheroidal harmonics exactly match that singularity.
The link with the image theory solutions can be seen more explicitly with the Havelock formula \cite{1952Havelock}, which for our offset spheroidal coordinate system can be re-expressed as:
\begin{align}
Q_n(\xib)P_n(\etab) = \frac{1}{2}\int_{0}^{R_i} \frac{P_n(2\tilde{z}/R_i-1)}{\sqrt{x^2+ y^2 + (z-\tilde z)^2}} d\tilde z.
\end{align}
The Legendre polynomials in the numerator are a basis for functions defined on the interval $0\leq\tilde{z}\leq R_i$. This makes $Q_n(\xib)P_n(\etab)$ a basis for any charge distribution on the segment OI, and the expansion will converge in all space (except the line segment).
A similar expression can be written for the radially-inverted offset spheroidal coordinate system:
\begin{align}
\frac{Q_n(\check{\xi})P_n(\check{\eta})}{r} = \frac{1}{2}\int_{R_e}^{\infty} \frac{P_n(2R_e/\tilde z-1)/\tilde{z}}{\sqrt{x^2+ y^2 + (z-\tilde z)^2}} d\tilde z.
\end{align}
so $Q_n(\check{\xi})P_n(\check{\eta})/r$ are natural functions for expressing semi-infinite line singularities.
These considerations may be fruitful in devising new approaches to improve the solutions of related problems of mathematical physics.


\vspace{2cm}
\appendix

\section{Relationships between spherical and spheroidal harmonics} \label{secRels}
In \cite{2017SuperLaplace} we presented two new expansions relating spherical and offset spheroidal harmonics in the case of
offset spheroidal coordinates defined for a general $c>0$ as:
\begin{align}
\xib_c &=\frac{r+r'_c}{c}, \quad \etab_c =\frac{r-r'_c}{c},  \label{EqOffsetSph}\\
\mathrm{where~} r'_c &= |{\b r} - c\unitz | = \sqrt{r^2-2cr\cos\theta + c^2}. \nonumber
\end{align}
In the text, those coordinates are used with $c\equiv R_i$ to express solutions outside the sphere.

We reproduce those expansions below for completeness and refer to \cite{2017SuperLaplace} for their proofs:
\begin{widetext}
\begin{align}
P_n^m(\xib_c)P_n^m(\etab_c)=&\frac{(n+m)!}{(n-m)!}\sum_{k=m}^n 
\frac{(-)^{n+k}}{k!(k+m)!}\frac{(n+k)!}{(n-k)!}\left(\frac{r}{c}\right)^kP_k^m(\cos\theta), \label{PPnmvsPnm} \\
\left(\frac{c}{r}\right)^{n+1}P_n^m(\cos\theta)&=\frac{2(-)^{n+m}}{n!(n-m)!}\sum_{k=n}^\infty (-)^k(2k+1)
\frac{(k+n)!}{(k-n)!}\frac{(k-m)!}{(k+m)!}Q_k^m(\xib_c)P_k^m(\etab_c). \label{PnmvsQPnm}
\end{align}

We also present and prove the inverse expansions, of which Eqs. \ref{PvsPP} and \ref{Qexpand} are special cases.
These are
%

\begin{align}
\left(\frac{r}{c}\right)^n P_n^m(\cos\theta)&=n!(n+m)!\sum_{k=m}^n~\frac{2k+1}{(n-k)!(n+k+1)!}\frac{(k-m)!}{(k+m)!} P_k^m(\bar\xi_c)P_k^m(\bar\eta_c) \label{PnmvsPPnm} 
\\
Q_n^m(\bar\xi_c)P_n^m(\bar\eta_c)&=\frac{(-)^m}{2}\frac{(n+m)!}{(n-m)!}\sum_{k=n}^\infty~\frac{k!(k-m)!}{(k-n)!(k+n+1)!} \left(\frac{c}{r}\right)^{k+1}P_k^m(\cos\theta). \label{QPnmvsPnm}  
\end{align}
\end{widetext}
Similar relationships have been given and proved in the case of the standard spheroidal coordinates centered at the origin \cite{1917HarmonicRelations,2000Jansen,2002Russian}.
Eq. \ref{PnmvsPPnm} is proved in the next section. Eq. \ref{QPnmvsPnm} can be proved following the same method used to prove Eq. \ref{PnmvsQPnm} in \cite{2017SuperLaplace} so the proof is not repeated here.

\section{Proof of Eq. \ref{PnmvsPPnm}}

Knowing that Eq.~\ref{PPnmvsPnm} holds and is a finite invertible basis transformation, it must be possible to find the inverse expansion:
\begin{align}
\left(\frac{r}{c}\right)^n P_n^m(\cos\theta)=\sum_{k=m}^n\alpha_{nk}^m P_k^m(\bar\xi)P_k^m(\bar\eta).
\end{align}
Substitute this into Eq. \ref{PPnmvsPnm}, and rearrange the order of summation to get
\begin{align*}
&P_n^m(\bar\xi)P_n^m(\bar\eta)=\nonumber\\
&\frac{(n+m)!}{(n-m)!}\sum_{p=m}^n\sum_{k=p}^n\frac{(-)^{n+k}}{k!(k+m)!}\frac{(n+k)!}{(n-k)!}  \alpha_{kp}^m P_p^m(\bar\xi)P_p^m(\bar\eta).
\end{align*}
Since $P_n^m(\bar\xi)P_n^m(\bar\eta)$ are orthogonal functions, we must have
\begin{align}
\frac{(n+m)!}{(n-m)!}\sum_{k=p}^n\frac{(-)^{n+k}}{k!(k+m)!}\frac{(n+k)!}{(n-k)!}  \alpha_{kp}^m = \delta_{np} \label{PnmvsPPnm alpha}.
\end{align}
The coefficients $\alpha_{kp}^m$ can be deduced by looking at the orthogonality relation for the shifted Legendre polynomials $P_n(2x-1)$:
\begin{align}
\int_0^1 P_n(2x-1)P_p(2x-1) \d x = \frac{\delta_{np}}{2p+1}.
\end{align}
Now expand the shifted Legendre polynomials in powers of $x$:
\begin{align}
&\frac{\delta_{np}}{2p+1} =\nonumber\\
&\int_0^1 \sum_{k=0}^n\frac{(-)^{n+k}(n+k)!}{k!^2(n-k)!}x^k \sum_{q=0}^p\frac{(-)^{q+p}(q+p)!}{q!^2(p-q)!}x^q \d x \nonumber\\
&= \sum_{k=0}^n\frac{(-)^{n+k}(n+k)!}{k!^2(n-k)!} \sum_{q=0}^p\frac{(-)^{q+p}(q+p)!}{q!^2(p-q)!} \frac{1}{k+q+1} \label{dnp}.
\end{align}
The sum over $q$ can be simplified by using Eq. \ref{EqnCn} with $n\rightarrow p$, $k\rightarrow q$, $\mu\rightarrow k+1$)
\begin{align}
\sum_{q=0}^p\frac{(-)^{q+p}(q+p)!}{q!^2(p-q)!} \frac{1}{k+q+1} = \frac{k!^2}{(k-p)!(k+p+1)!}.
\end{align}
Substituting this back into Eq. \ref{dnp} we have
\begin{align}
\sum_{k=0}^n \frac{(-)^{n+k}(n+k)!}{(n-k)!(k-p)!(k+p+1)!}=\frac{\delta_{np}}{2p+1}.
\end{align}
Compare this with Eq. \ref{PnmvsPPnm alpha} to find that $\alpha_{nk}^m$ are in fact the coefficients given in Eq. \ref{PnmvsPPnm}. \ensuremath{\Box}\\

\begin{figure}[b]
\includegraphics[scale=.65]{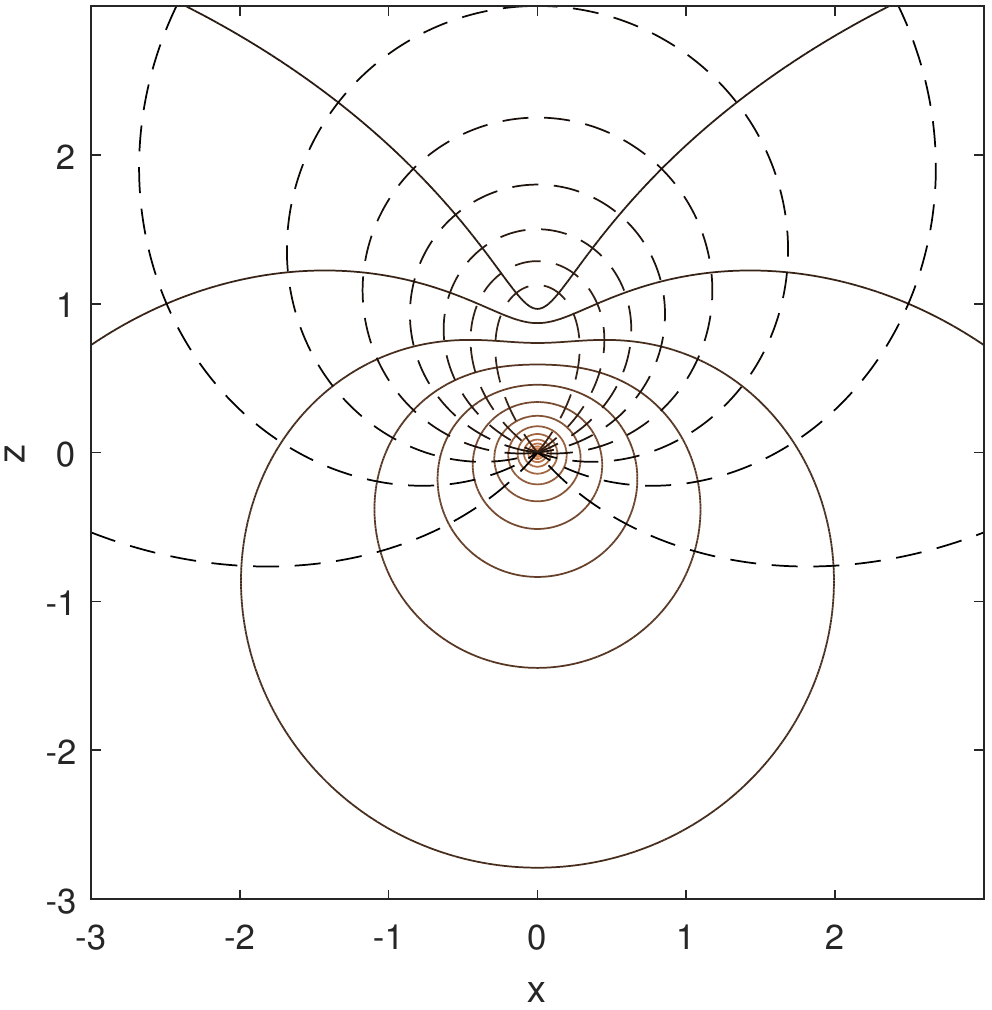}
\caption{Constant coordinate surfaces on the plane $y=0$. Surfaces of constant $\check{\eta}$ (dashed lines) range from a finite line on the $z$-axis for $\check{\eta}=1$ to sharply dimpled spheres for $\check{\eta}<0$. $\check{\xi}$ range from small spheres for large $\check{\xi}$ to smoothly dimpled spheres as $\check{\xi}\rightarrow1$. The focal length of the coordinates is 1.}
\label{FigIsoCoords}
\end{figure}
\begin{figure*}
\includegraphics[scale=.6]{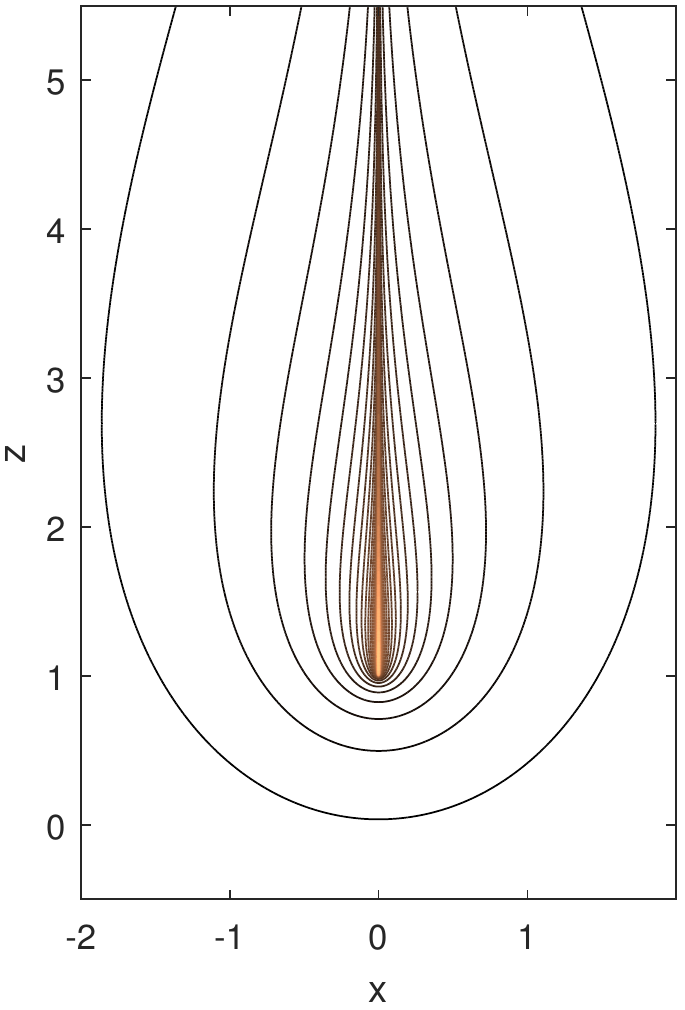}
\includegraphics[scale=.6]{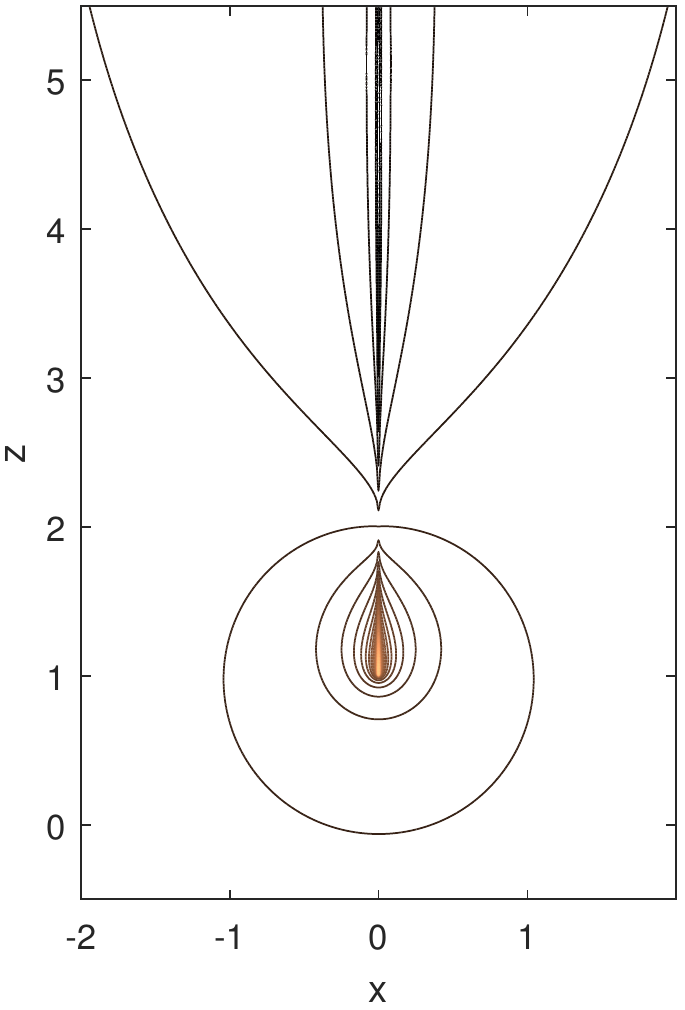}
\includegraphics[scale=.6]{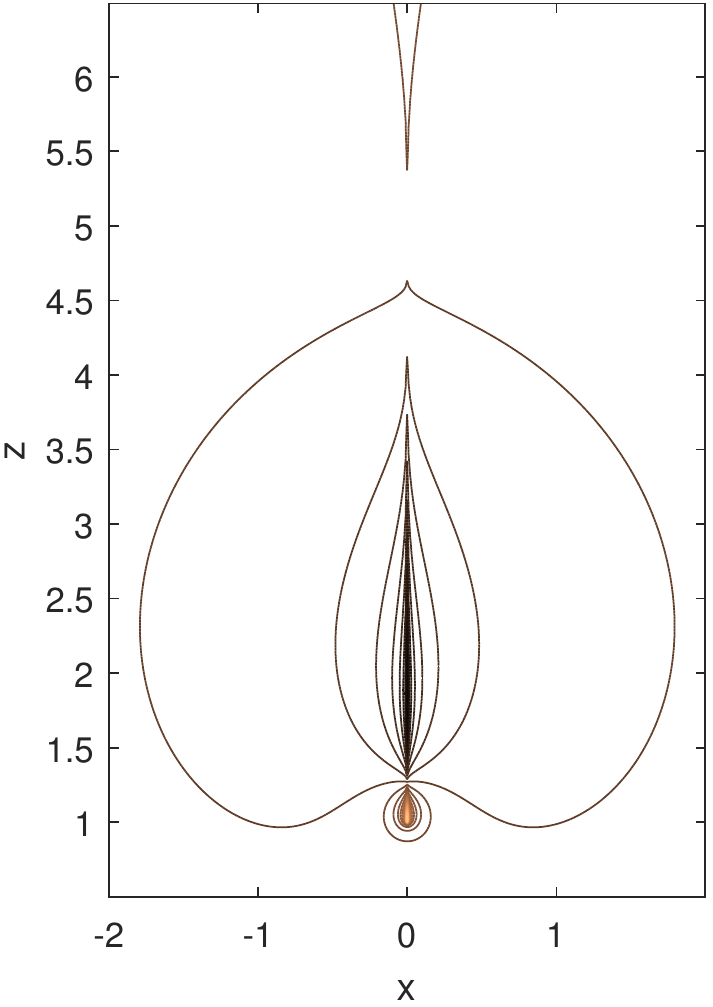}
\includegraphics[scale=.6]{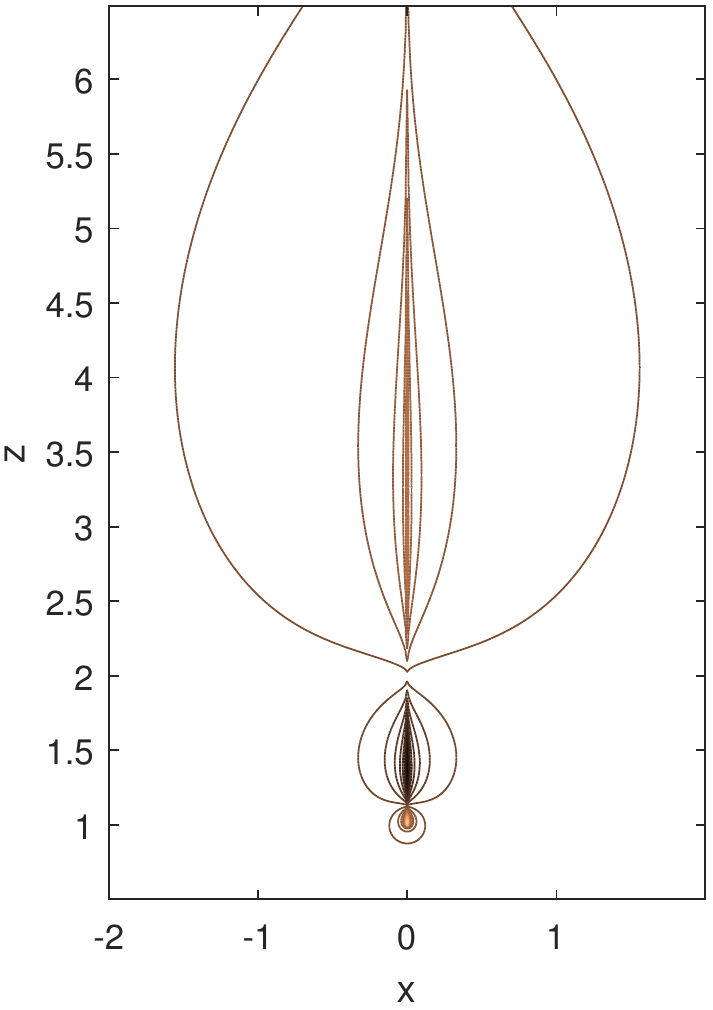}
\caption{Isopotentials of the functions $Q_n(\check{\xi})P_n(\check{\eta})/r$ for $n=0, 1, 2, 3$ from left to right. The focal length of the non-inverted spheroidal coordinates is 1, so that they are singular on the $z$-axis from $z=1$ to $\infty$.}
\label{FigIsoV}
\end{figure*}

\section{Boundary of convergence of Eq.~\ref{VPPF}}\label{convProof}

Here we prove that the boundary of convergence of Eq. \ref{VPPF} is a spheroid whose surface touches point E. To do this we look at the limit as $n\rightarrow\infty$ of the terms in the series. The asymptotic form of the Legendre polynomials $P_n(u)$ for $|u|<1$ is \citep{1931Hobson} (pp 304-306):
\begin{align}
P_n(u)\rightarrow\sqrt{\frac{2}{\pi n\sqrt{1-u^2}}}\sin\left(\left(n+\frac{1}{2}\right)\cos^{-1}u+\frac{\pi}{4}\right),
\end{align}
which applies to $P_n(\bar\eta_f)$. For $P_n(\bar\xi_f)$, we need the asymptotic form for $|u|>1$:
\begin{align}
P_n(u) \rightarrow \frac{(u+\sqrt{u^2-1})^{n+1/2}}{\sqrt{2\pi n}(u^2-1)^{1/4}}.
\end{align} 
To determine the asymptotic form of the sum over $k$, we evaluate Eq. \ref{QPnmvsPnm} (with $c\rightarrow f$) at $\cos\theta=\eta_f=1$:
\begin{align}
\sum_{k=n}^\infty\frac{k!^2}{(k-n)!(k+n+1)!}\left(\frac{f}{r}\right)^{k+1}=2Q_n(\bar\xi_f). \label{Qexpand}
\end{align}
Integrating this expression with respect to $v=\bar\xi_f=2r/f-1$ from $r=\infty$ to $r=R_e$:
\begin{align}
\sum_{k=n}^\infty\frac{k!^2(f/R_e)^k}{(k-n)!(k+n+1)!}\frac{-1}{k}=\int_\infty^{2R_e/f-1}Q_n(v) \d v.
\end{align}
As $n\rightarrow\infty$, the left hand side is proportional to the sum over $k$ in Eq. \ref{VPPF}. Evaluating the right hand side:
\begin{align}
\int^u Q_n(v) \d v = \frac{u Q_n(u)-Q_{n+1}(u)}{n+1}.
\end{align}
The asymptotic form of $Q_n(u)$ for $|u|>1$ is \citep{1931Hobson} 
\begin{align}
Q_n(u) \rightarrow \frac{\sqrt{\pi/2/n}}{(u^2-1)^{1/4}(u+\sqrt{u^2-1})^{n+1/2}}, 
\end{align}
so that
\begin{align}
\int_\infty^u Q_n(v)\d v \rightarrow  \frac{\mathrm{constant}}{n^{3/2}(u+\sqrt{u^2-1})^n}. \label{limQn}
\end{align}
Putting all this together, the $n^{th}$ term in the series in Eq.~\ref{VPP} approaches
\begin{align}
\frac{\mathrm{constant}}{n}\left(\frac{\bar\xi_f+\sqrt{\bar\xi_f^2-1}}{2R_e/f-1+\sqrt{(2R_e/f-1)^2-1}}\right)^n\! P_n(\etab).
\end{align}
Let $X$ be the expression in the large brackets. If $X<1$ (equivalent to $\bar\xi_f<2r/f-1$) it is clear that the series converges since it is bounded by the series $\sum X^n/n$ which converges. For $X>1$ ($\bar\xi_f>2R_e/f-1$), the series diverges because the terms increase in size. Geometrically, the boundary of convergence is the surface of a spheroid with foci at $z=0$ and $z=f$ that passes through point E at $R_e\hat{\b z}$.

\section{Inverted spheroidal coordinate surfaces}
\label{SecInvSph}
For insight we plot the surfaces of constant $\check{\xi}$ and $\check{\eta}$ in Fig.~\ref{FigIsoCoords}, and equipotentials of the first few orders of the harmonics $Q_n(\check{\xi})P_n(\check{\eta})/r$ in Fig.~\ref{FigIsoV}.

\section*{Acknowledgements}
This work was supported by the MacDiarmid Institute of advanced materials and nanotechnology, and a Victoria doctoral scholarship.\\

\bibliography{Tmatrix}

\begin{thebibliography}{18}%
\makeatletter
\providecommand \@ifxundefined [1]{%
 \@ifx{#1\undefined}
}%
\providecommand \@ifnum [1]{%
 \ifnum #1\expandafter \@firstoftwo
 \else \expandafter \@secondoftwo
 \fi
}%
\providecommand \@ifx [1]{%
 \ifx #1\expandafter \@firstoftwo
 \else \expandafter \@secondoftwo
 \fi
}%
\providecommand \natexlab [1]{#1}%
\providecommand \enquote  [1]{``#1''}%
\providecommand \bibnamefont  [1]{#1}%
\providecommand \bibfnamefont [1]{#1}%
\providecommand \citenamefont [1]{#1}%
\providecommand \href@noop [0]{\@secondoftwo}%
\providecommand \href [0]{\begingroup \@sanitize@url \@href}%
\providecommand \@href[1]{\@@startlink{#1}\@@href}%
\providecommand \@@href[1]{\endgroup#1\@@endlink}%
\providecommand \@sanitize@url [0]{\catcode `\\12\catcode `\$12\catcode
  `\&12\catcode `\#12\catcode `\^12\catcode `\_12\catcode `\%12\relax}%
\providecommand \@@startlink[1]{}%
\providecommand \@@endlink[0]{}%
\providecommand \url  [0]{\begingroup\@sanitize@url \@url }%
\providecommand \@url [1]{\endgroup\@href {#1}{\urlprefix }}%
\providecommand \urlprefix  [0]{URL }%
\providecommand \Eprint [0]{\href }%
\providecommand \doibase [0]{http://dx.doi.org/}%
\providecommand \selectlanguage [0]{\@gobble}%
\providecommand \bibinfo  [0]{\@secondoftwo}%
\providecommand \bibfield  [0]{\@secondoftwo}%
\providecommand \translation [1]{[#1]}%
\providecommand \BibitemOpen [0]{}%
\providecommand \bibitemStop [0]{}%
\providecommand \bibitemNoStop [0]{.\EOS\space}%
\providecommand \EOS [0]{\spacefactor3000\relax}%
\providecommand \BibitemShut  [1]{\csname bibitem#1\endcsname}%
\let\auto@bib@innerbib\@empty
\bibitem [{\citenamefont {Maji\ifmmode~\acute{c}\else \'{c}\fi{}}\ \emph
  {et~al.}(2017)\citenamefont {Maji\ifmmode~\acute{c}\else \'{c}\fi{}},
  \citenamefont {Augui\'e},\ and\ \citenamefont {Le~Ru}}]{2017SuperLaplace}%
  \BibitemOpen
  \bibfield  {author} {\bibinfo {author} {\bibfnamefont {M.~R.~A.}\
  \bibnamefont {Maji\ifmmode~\acute{c}\else \'{c}\fi{}}}, \bibinfo {author}
  {\bibfnamefont {B.}~\bibnamefont {Augui\'e}}, \ and\ \bibinfo {author}
  {\bibfnamefont {E.~C.}\ \bibnamefont {Le~Ru}},\ }\href {\doibase
  10.1103/PhysRevE.95.033307} {\bibfield  {journal} {\bibinfo  {journal} {Phys.
  Rev. E}\ }\textbf {\bibinfo {volume} {95}},\ \bibinfo {pages} {033307}
  (\bibinfo {year} {2017})}\BibitemShut {NoStop}%
\bibitem [{\citenamefont {Thomson}(1847)}]{1847Thomson}%
  \BibitemOpen
  \bibfield  {author} {\bibinfo {author} {\bibfnamefont {W.}~\bibnamefont
  {Thomson}},\ }\href {http://eudml.org/doc/233970} {\bibfield  {journal}
  {\bibinfo  {journal} {Journal de Math\'{e}matiques Pures et Appliqu\'{e}es}\
  }\textbf {\bibinfo {volume} {12}},\ \bibinfo {pages} {256} (\bibinfo {year}
  {1847})}\BibitemShut {NoStop}%
\bibitem [{\citenamefont {Dassios}(2009)}]{2009Dassios}%
  \BibitemOpen
  \bibfield  {author} {\bibinfo {author} {\bibfnamefont {G.}~\bibnamefont
  {Dassios}},\ }\href@noop {} {\bibfield  {journal} {\bibinfo  {journal} {IMA
  J. Appl. Math.}\ }\textbf {\bibinfo {volume} {74}},\ \bibinfo {pages} {427}
  (\bibinfo {year} {2009})}\BibitemShut {NoStop}%
\bibitem [{\citenamefont {Amaral}\ \emph {et~al.}(2017)\citenamefont {Amaral},
  \citenamefont {Ventura},\ and\ \citenamefont {Lemos}}]{2017Kelvin}%
  \BibitemOpen
  \bibfield  {author} {\bibinfo {author} {\bibfnamefont {R.}~\bibnamefont
  {Amaral}}, \bibinfo {author} {\bibfnamefont {O.}~\bibnamefont {Ventura}}, \
  and\ \bibinfo {author} {\bibfnamefont {N.}~\bibnamefont {Lemos}},\
  }\href@noop {} {\bibfield  {journal} {\bibinfo  {journal} {European Journal
  of Physics}\ }\textbf {\bibinfo {volume} {38}},\ \bibinfo {pages} {025206}
  (\bibinfo {year} {2017})}\BibitemShut {NoStop}%
\bibitem [{\citenamefont {Stratton}(1941)}]{1941Stratton}%
  \BibitemOpen
  \bibfield  {author} {\bibinfo {author} {\bibfnamefont {J.~A.}\ \bibnamefont
  {Stratton}},\ }\href@noop {} {\emph {\bibinfo {title} {Electromagnetic
  theory}}}\ (\bibinfo  {publisher} {McGraw-Hill},\ \bibinfo {address} {New
  York},\ \bibinfo {year} {1941})\BibitemShut {NoStop}%
\bibitem [{\citenamefont {Neumann}(1883)}]{1883Neumann}%
  \BibitemOpen
  \bibfield  {author} {\bibinfo {author} {\bibfnamefont {C.}~\bibnamefont
  {Neumann}},\ }\href@noop {} {\emph {\bibinfo {title} {Hydrodynamische
  Untersuchungen nebst einem anhange {\"u}ber die Probleme der Elektrostatik
  und der Magnetischen Induction}}}\ (\bibinfo  {publisher} {Teubner},\
  \bibinfo {address} {Leipzig},\ \bibinfo {year} {1883})\BibitemShut {NoStop}%
\bibitem [{\citenamefont {Poladian}(1988)}]{poladian1988}%
  \BibitemOpen
  \bibfield  {author} {\bibinfo {author} {\bibfnamefont {L.}~\bibnamefont
  {Poladian}},\ }\href@noop {} {\bibfield  {journal} {\bibinfo  {journal}
  {Quart. J. Mech. Appl. Math.}\ }\textbf {\bibinfo {volume} {41}},\ \bibinfo
  {pages} {395} (\bibinfo {year} {1988})}\BibitemShut {NoStop}%
\bibitem [{\citenamefont {Lindell}(1992)}]{1992Lindell}%
  \BibitemOpen
  \bibfield  {author} {\bibinfo {author} {\bibfnamefont {I.~V.}\ \bibnamefont
  {Lindell}},\ }\href@noop {} {\bibfield  {journal} {\bibinfo  {journal} {Radio
  Science}\ }\textbf {\bibinfo {volume} {27}},\ \bibinfo {pages} {1} (\bibinfo
  {year} {1992})}\BibitemShut {NoStop}%
\bibitem [{\citenamefont {Havelock}(1952)}]{1952Havelock}%
  \BibitemOpen
  \bibfield  {author} {\bibinfo {author} {\bibfnamefont {T.~H.}\ \bibnamefont
  {Havelock}},\ }\href@noop {} {\bibfield  {journal} {\bibinfo  {journal}
  {Quart. J. Mech. Appl. Math.}\ }\textbf {\bibinfo {volume} {5}},\ \bibinfo
  {pages} {129} (\bibinfo {year} {1952})}\BibitemShut {NoStop}%
\bibitem [{\citenamefont {Miloh}(1974)}]{1974Miloh}%
  \BibitemOpen
  \bibfield  {author} {\bibinfo {author} {\bibfnamefont {T.}~\bibnamefont
  {Miloh}},\ }\href@noop {} {\bibfield  {journal} {\bibinfo  {journal} {SIAM J.
  Appl. Math.}\ }\textbf {\bibinfo {volume} {26}},\ \bibinfo {pages} {334}
  (\bibinfo {year} {1974})}\BibitemShut {NoStop}%
\bibitem [{\citenamefont {Parry~Moon}(1961)}]{1961FieldTheoryHandbook}%
  \BibitemOpen
  \bibfield  {author} {\bibinfo {author} {\bibfnamefont {D.~E. S.~a.}\
  \bibnamefont {Parry~Moon}},\ }\href
  {http://gen.lib.rus.ec/book/index.php?md5=B558530211DD73F2D830B2530CF83257}
  {\emph {\bibinfo {title} {Field Theory Handbook: Including Coordinate
  Systems, Differential Equations and Their Solutions}}}\ (\bibinfo
  {publisher} {Springer Berlin Heidelberg},\ \bibinfo {year}
  {1961})\BibitemShut {NoStop}%
\bibitem [{\citenamefont {Dassios}\ and\ \citenamefont
  {Miloh}(1999)}]{dassios1999rayleigh}%
  \BibitemOpen
  \bibfield  {author} {\bibinfo {author} {\bibfnamefont {G.}~\bibnamefont
  {Dassios}}\ and\ \bibinfo {author} {\bibfnamefont {T.}~\bibnamefont
  {Miloh}},\ }\href@noop {} {\bibfield  {journal} {\bibinfo  {journal}
  {Quarterly of Applied Mathematics}\ }\textbf {\bibinfo {volume} {57}},\
  \bibinfo {pages} {757} (\bibinfo {year} {1999})}\BibitemShut {NoStop}%
\bibitem [{\citenamefont {Hadjinicolaou}\ and\ \citenamefont
  {Protopapas}(2015)}]{2015InvertedSpheroidal}%
  \BibitemOpen
  \bibfield  {author} {\bibinfo {author} {\bibfnamefont {M.}~\bibnamefont
  {Hadjinicolaou}}\ and\ \bibinfo {author} {\bibfnamefont {E.}~\bibnamefont
  {Protopapas}},\ }\href@noop {} {\bibfield  {journal} {\bibinfo  {journal}
  {IMA J. Appl. Math.}\ }\textbf {\bibinfo {volume} {80}},\ \bibinfo {pages}
  {1475} (\bibinfo {year} {2015})}\BibitemShut {NoStop}%
\bibitem [{\citenamefont {Sten}\ and\ \citenamefont
  {Lindell}(1992)}]{1992LindellInside}%
  \BibitemOpen
  \bibfield  {author} {\bibinfo {author} {\bibfnamefont {J.~C.-E.}\
  \bibnamefont {Sten}}\ and\ \bibinfo {author} {\bibfnamefont {I.~V.}\
  \bibnamefont {Lindell}},\ }\href@noop {} {\bibfield  {journal} {\bibinfo
  {journal} {Microwave and optical technology letters}\ }\textbf {\bibinfo
  {volume} {5}},\ \bibinfo {pages} {597} (\bibinfo {year} {1992})}\BibitemShut
  {NoStop}%
\bibitem [{\citenamefont {Jeffery}(1917)}]{1917HarmonicRelations}%
  \BibitemOpen
  \bibfield  {author} {\bibinfo {author} {\bibfnamefont {G.~B.}\ \bibnamefont
  {Jeffery}},\ }\href {\doibase 10.1112/plms/s2-16.1.133} {\bibfield  {journal}
  {\bibinfo  {journal} {Proceedings of the London Mathematical Society}\
  }\textbf {\bibinfo {volume} {s2-16}},\ \bibinfo {pages} {133} (\bibinfo
  {year} {1917})}\BibitemShut {NoStop}%
\bibitem [{\citenamefont {Jansen}(2000)}]{2000Jansen}%
  \BibitemOpen
  \bibfield  {author} {\bibinfo {author} {\bibfnamefont {G.}~\bibnamefont
  {Jansen}},\ }\href@noop {} {\bibfield  {journal} {\bibinfo  {journal} {J.
  Phys. A: Math. and General}\ }\textbf {\bibinfo {volume} {33}},\ \bibinfo
  {pages} {1375} (\bibinfo {year} {2000})}\BibitemShut {NoStop}%
\bibitem [{\citenamefont {Antonov}\ and\ \citenamefont
  {Baranov}(2002)}]{2002Russian}%
  \BibitemOpen
  \bibfield  {author} {\bibinfo {author} {\bibfnamefont {V.~A.}\ \bibnamefont
  {Antonov}}\ and\ \bibinfo {author} {\bibfnamefont {A.~S.}\ \bibnamefont
  {Baranov}},\ }\href@noop {} {\bibfield  {journal} {\bibinfo  {journal}
  {Technical Physics}\ }\textbf {\bibinfo {volume} {47}},\ \bibinfo {pages}
  {80} (\bibinfo {year} {2002})}\BibitemShut {NoStop}%
\bibitem [{\citenamefont {Hobson}(1931)}]{1931Hobson}%
  \BibitemOpen
  \bibfield  {author} {\bibinfo {author} {\bibfnamefont {E.~W.}\ \bibnamefont
  {Hobson}},\ }\href@noop {} {\emph {\bibinfo {title} {The theory of spherical
  and ellipsoidal harmonics}}}\ (\bibinfo  {publisher} {CUP Archive},\ \bibinfo
  {year} {1931})\BibitemShut {NoStop}%
\end{thebibliography}%

\end{document}